\documentclass[conference]{IEEEtran}
\IEEEoverridecommandlockouts
\usepackage{cite}
\usepackage{amsmath,amssymb,amsfonts}
\usepackage{algpseudocode}

\usepackage{graphicx,epsfig,epstopdf}
\usepackage{textcomp}
\usepackage{xcolor}
\def\BibTeX{{\rm B\kern-.05em{\sc i\kern-.025em b}\kern-.08em
    T\kern-.1667em\lower.7ex\hbox{E}\kern-.125emX}}
\begin{document}

\title{Centralized and distributed schedulers for non-coherent joint transmission
}

\author{\IEEEauthorblockN{ Shangbin~Wu and Yinan~Qi}
\IEEEauthorblockA{\textit{Samsung R\&D Institute UK} \\
Staines-upon-Thames, United Kingdom \\
\{shangbin.wu, yinan.qi\}@samsung.com}
}

\maketitle

\begin{abstract}
This paper studies the performance of three typical network coordination schemes, i.e., dynamic point selection, fully overlapped non-coherent joint transmission (F-NCJT), and non-fully overlapped NCJT (NF-NCJT), in 3GPP new radio (NR) in indoor scenarios via system level simulation. Each of these schemes requires a different level of user data and channel state information (CSI) report exchange among coordinated transmission reception points (TRPs) depending on centralized or distributed schedulers. Scheduling strategies of these network coordination schemes are briefly discussed. It has been demonstrated that distributed network coordination schemes (e.g., NF-NCJT) can still perform reasonably well; a result which has important implications to the design of the fifth generation (5G) cellular network architecture.
\end{abstract}


\section{Introduction}
The demand for high data rate, energy efficient, and robust communications with significantly improved user experience drives the development of the fifth generation (5G) cellular networks. The data rate of 5G is expected to achieve tens of times larger than the legacy long term evolution advanced (LTE-A) system \cite{Andrews14}\cite{38913}. Besides, latency and mobility interruption time should be improved to guarantee smooth transitions when users are traveling from cell to cell. In 5G, quality of experience (QoE) of cell edge users needs to be largely enhanced. 

Traditionally, cellular mobile communication networks are built upon the concept of a mobile terminal being connected to a single serving base station (BS). Each BS in legacy networks focused on allocating resources optimally to its own attached user equipments (UEs). However, with denser and heterogeneous deployments employed, focus has shifted away from this single serving cell paradigm. Network coordination concepts such as coordinated multipoint (CoMP) \cite{Lee12} transmission schemes are now being leveraged to push greater data rates and see higher spectral efficiency gains. CoMP schemes use multiple transmission reception points (TRPs) in a coordinated manner by allowing TRPs to mutually exchange information. In this case, signals from coordinated TRPs can be suppressed, or, used constructively.

The development of CoMP in the $3^{\mathrm{rd}}$ generation partnership project (3GPP) can be traced back to LTE-A release 11 (Rel--11), where several downlink CoMP schemes such as joint transmission (JT) \cite{Davydov13}, dynamic point selection (DPS) \cite{Agrawal14}, dynamic point blanking (DPB) and coordinated scheduling/beamforming (CS/CB) \cite{Sun12} were investigated and evaluated. The JT category can be further divided into two groups, i.e., coherent JT (CJT) and non-coherent JT (NCJT). CJT performs joint beamforming from all coordinated TRPs, which can be regarded as a distributed multiple-input multiple-output (MIMO) system. Contrarily, NCJT allows coordinated TRPs to transmit independent layers to the target UE.

In LTE-A Rel-11, major features pertinent to CoMP operation including support of channel state information (CSI) feedback with multiple CSI processes, virtual cell identity, and quasi-co-location (QCL) support to RSs are standardized. Two QCL antenna ports are assumed to have similar delay spreads, Doppler spreads, delays, and channel gains. A new transmission mode (TM) 10 for supporting CoMP operations has been specified. TM10 is similar to TM9 to support up to 8 layers and it supports configuring UEs two different QCL types (Type A or Type B) by radio resource control (RRC) signalling in the current specification. Type A essentially implies all reference signal (RS) ports configured to the UE are QCL, and Type B lets the UE perform DPS for physical downlink chared channel (PDSCH) reception by indicating that one CSI-RS resource is QCL with the PDSCH demodulation RS (DMRS) in a given subframe. In the later release, e.g., Rel-12, CoMP for non-ideal backhaul link case was addressed and enhanced CoMP considering inter-eNB CoMP was standardized.

Recently, the study item (SI) on further enhancements on CoMP (FeCoMP) in Rel-14 has finished and a follow-up work item (WI) has started in Rel-15 \cite{RP170750}\cite{36741}. In the SI, both NCJT and CS/CB with full dimension MIMO (FD-MIMO) are investigated and evaluated but only NCJT is chosen in the WI for standardization because of more significant performance gains, especially in the indoor scenario. The development of CoMP continues in the 5G new radio (NR) where both CJT and NCJT will be studied and further standardization enhancements are expected to fully harvest the gain of CoMP operations.

The approaches in each CoMP category present unique backhaul demands and implementation complexities. Since CJT performs joint beamforming, it requires backhaul links with high capacity and low latency as well as synchronization among coordinated TRPs. Yet another reason NCJT is of considerable interest is that it only requires little data exchange among TRPs unlike other CoMP techniques. NCJT operation handles each transmission from a TRP to UE individually. This means scheduling, rank and precoding matrix selection, and modulation and coding scheme (MCS) selection may be made individually per TRP. NCJT is often further categorized into fully overlapped NCJT (F-NCJT) and non-fully overlapped NCJT (NF-NCJT), respectively. F-NCJT requires that resources are allocated on equivalent physical resource blocks (PRBs) on each coordinated TRP, where as NF-NCJT sees further decoupling of the TRPs, allowing for flexible allocation across the available PRBs of each. A brief comparison between different network coordination schemes is listed in Table \ref{tab_NC_comparison}.

\begin{table}[ht]
\caption{Comparison between different network coordination schemes.}
\center

    \begin{tabular}{|c|c|c|c|c|}
    \hline
      & CJT & DPS & F-NCJT & NF-NCJT  \\ \hline
Centralized scheduler  & \checkmark & \checkmark & \checkmark & $\times$   \\ \hline        
        Joint precoder  & \checkmark   & $\times$   & $\times$ & $\times$  \\ \hline
        
        Fully overlapped PRB  & \checkmark  & $\times$   & \checkmark & $\times$   \\ \hline
        User data sharing & \checkmark & \checkmark & \checkmark & $\times$  \\ \hline

    \end{tabular}
    \label{tab_NC_comparison}
\end{table}

\begin{figure}[t]
	\centering
	\includegraphics[width=3.5in]{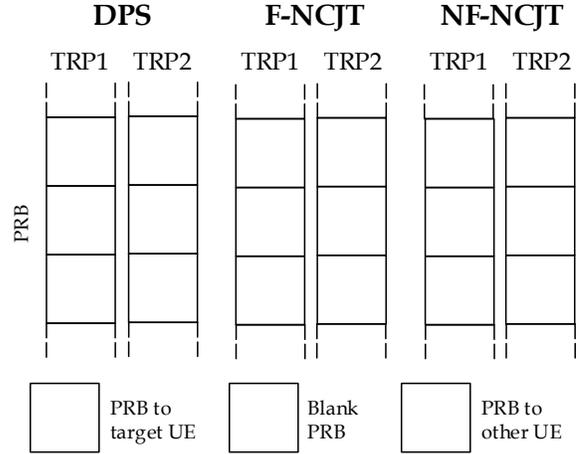}    
	\caption{Comparative Illustration of PRB allocation between schemes}
	\label{fig_comparePRBalloc}
\end{figure}

Examples of PRB allocations of DPS, F-NCJT, and NF-NCJT are illustrated in Fig.~\ref{fig_comparePRBalloc} with two TRPs, but the concept is extendible to multiple TRPs. The DPS technique serves as a CoMP baseline. Users will dynamically switch between TRPs with the goal of being served by the strongest TRP during each time interval. With DPS, the users is served from only one TRP in any time. It is possible within NF-NCJT for PRB allocation to partially overlap or, with greater improbability even fully overlap. The importance of this further distinction between F-NCJT and NF-NCJT schemes lies in the consequences each allocation scheme presents with regard to resource scheduling as well as radio access network (RAN) design. F-NCJT is suitable in the cloud RAN (CRAN) \cite{Wu15} architecture which assumes that a central processing unit exists to perform all resource allocations. Conversely, NF-NCJT fits the distributed RAN (DRAN) \cite{Awan16} architecture where local schedulers are embedded in TRPs. Given the use of NF-NCJT allows each TRP to schedule completely independently, this paper compares a centralized scheduling approach with distributed, per TRP scheduling. Based on this analysis, recommendations are made with respect to the potential throughput benefit versus the complexity cost of a centralized versus distributed approach.

The remainder of this paper is organized as follows. Section~\ref{sec_Network_coordination} outlines the system model used and mathematical notations carried forward for F-NCJT, NF-NCJT and DPS. This is followed in Section~\ref{sec_results_and_analysis} by results and analysis of system level simulations conducted to compare each scheme. Furthermore, the effect of increasing the number of cooperating nodes is simulated and results analysed. Conclusions are drawn in Section~\ref{conclusion_section}. 

\section{Network coordination schemes} \label{sec_Network_coordination}
\subsection{System model and notations}
Let us consider a multi-TRP network with $N_b$ TRPs. Each TRP is equipped with $N_T$ transmit antennas and each UE is equipped with $N_R$ receive antennas. For simplicity, each TRP will perform single layer transmission to a UE and these TRPs are grouped into a number of disjoint CoMP sets, i.e., a TRP will not be a member of two different CoMP sets. Let $\mathcal{C}_b$ denote the CoMP set that the $b$th ($b=1,2,\cdots, N_b$) TRP belongs to and the TRP index set in $\mathcal{C}_b$ be $\left\lbrace b_1,b_2,\cdots,b_{|\mathcal{C}_b|}\right\rbrace$. Also, let $U_b$ be the user vector of the $b$th TRP. TRPs are interconnected via backhaul and allowed to exchange certain information depending on the network coordination schemes. In this paper, latency of backhaul is ignored.

\begin{figure}[t]
\centering
\includegraphics[width=3.5in]{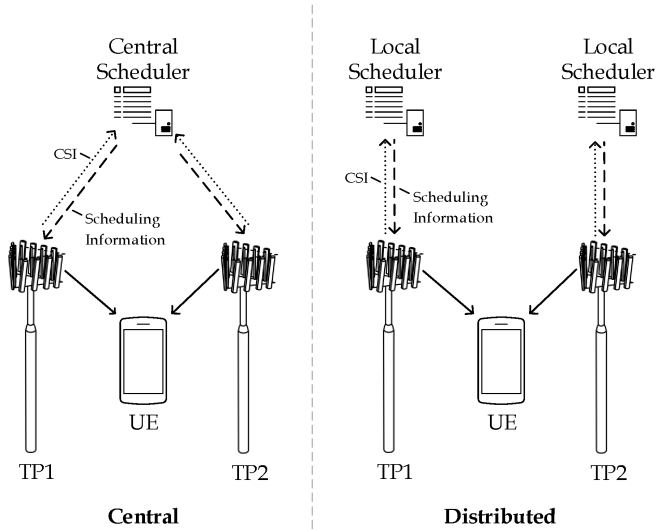}    
\caption{Diagram of centralized and distributed schedulers.}
\label{fig_CentralVDist_RA}
\end{figure}

Regarding resource scheduling approaches, as depicted in Fig. \ref{fig_CentralVDist_RA}, in the centralized approach, a centralized scheduler will collect all CSI reports from TRPs in a CoMP set and perform resource scheduling. Then, the centralized scheduler will send scheduling information back to TRPs. A centralized approach requires signalling exchange over the backhaul. As a result, it has higher capacity and latency requirements on backhaul links. On the hand, in the distributed approach, each TRP in a CoMP set will have a local scheduler, which collects CSI report from the UE. Only limited or no information will be exchanged further among TRPs. Hence, distributed approaches are more tolerant to non-ideal backhauls.

Each UE has a serving TRP (say the $b$th TRP) and will feedback CSI measurement reports to TRPs in $\mathcal{C}_b$. Once the CSI report from the $j$th UE in $U_b$ to the TRP with index $b_i$ in $\mathcal{C}_b$ is known, its individual spectral efficiency $c_{j,b_i}$ can be determined. Also, minimum mean squared error interference rejection combining (MMSE-IRC) receiver \cite{Tavares14} is assumed at the UE side. To perform MMSE-IRC, a UE needs to estimate the desired channel matrix and the covariance matrix of the interference channel. This can be done in the RS and CSI-RS framework in LTE or NR \cite{36213}. The serving TRP can allocate DMRS to a UE for channel estimation and zero power (ZP) CSI-RS for interference measurement. A tradeoff exists between the resources for interference measurement and throughput. More ZP CSI-RS resources can yield more accurate estimation of interference. However, these will reduce resources for useful signal transmission.
 
In the following paragraphs, for notation convenience, equations are shown on a per PRB basis. Hence, the  subscript denoting PRB index is dropped. Also, the target UE is assumed to the UE $k$, whose serving TRP is TRP $b$.

\subsection{DPS}
DPS schedules the UE in the TRP with the highest metric (e.g., received signal power) while blanking other coordinated TRPs to mitigate inter-TRP interference. As a result, let $\mathbf{I}_o$ be the interference from TRPs outside $\mathcal{C}_b$ and $\mathbf{n}$ be the addictive white Gaussian noise (AWGN) vector, the received signal $\mathbf{y}_k$ can be presented as  

\begin{align}
\mathbf{y}_k=\mathbf{H}_k\begin{bmatrix}
\mathbf{W}_{k,1} &  &  & \\ 
 & \mathbf{0}_{N_T\times 1} &  & \\ 
 &  & \ddots & \\ 
 &  &  & \mathbf{0}_{N_T\times 1}
\end{bmatrix}\mathbf{s}_k+\mathbf{I}_o+\mathbf{n}.
\end{align}
where $\mathbf{H}_k$ is the $N_R\times N_T|\mathcal{C}_b|$ aggregate channel matrix from coordinated TRPs in $\mathcal{C}_b$. Here, for notation convenience, it is assumed that the UE is scheduled by the first TRP within the CoMP set. The $N_T\times 1 $ matrix $\mathbf{W}_{k,1}$ denotes the $k$th user's precoding matrix from the first coordinated TRP in $\mathcal{C}_b$. The TRP index can be modified accordingly. In this case, the symbol vector $\mathbf{s}_k$ consists of the first symbol for the user and $|\mathcal{C}_b|-1$ zero entries, i.e.,  
$\mathbf{s}_k=[{s}_{k,1};\mathbf{0}_{(|\mathcal{C}_b|-1)\times 1}]^T$.

The scheduling algorithm of DPS is described in Fig. \ref{fig_DPS}. First, all UEs within a CoMP set need to be added to the user vector of each coordinated TRP. Second, the maximum  spectral efficiency of each UE among all TRPs is calculated as inputs to the centralized scheduler. The centralized scheduler then allocates resources to UEs according to the proportional fair rule. The whole scheduling procedure needs to be done only once for each CoMP set. Each TRP simply reads the scheduling information after the scheduling procedure. It should be noted that both user data and CSI reports need to be fully shared among all coordinated TRPs in a CoMP set.   

\begin{figure}
\hrulefill
\begin{algorithmic}[1]
\For{$b=1,2,\cdots, N_b$}
\State User vector $U_b$ of the $b$th TRP
\State $U'_b=U_b$
\State Identify the CoMP set $\mathcal{C}_b$ of the $b$th TRP
\State Read the indices $\left\lbrace b_1,b_2,\cdots,b_{|\mathcal{C}_b|}\right\rbrace$ of TRPs in $\mathcal{C}_b$
\For{$i=1,2,\cdots,|\mathcal{C}_b|$}
\State $U_b=U_b\bigcup U_{b_i}$
\EndFor
\If{$\mathcal{C}_b$ is scheduled}
\State The $b$th TRP reads PRB allocations from $\mathcal{C}_b$
\State Blank PRBs occupied by users not in $U'_b$
\Else
\For {$j=1,2,\cdots,|U_b|$}
\State Compute aggregate spectral efficiency $\eta_j=\max \{c_{j,b_1},c_{j,b_2},\cdots,c_{j,b_{|\mathcal{C}_b|}} \}  $
\EndFor
\State Perform proportional fair scheduling for $\mathcal{C}_b$
\State $\mathcal{C}_b$ is scheduled = true
\EndIf
\EndFor
\end{algorithmic}
\hrulefill
\caption{Scheduling algorithm for DPS.}
\label{fig_DPS}
\end{figure}

The main gain of DPS is from the blank PRBs from neighbor coordinated TRPs, which can reduce interference to cell edge users.

\subsection{F-NCJT}
In F-NCJT, coordinated TRPs transmit on the same PRB to a user but with different codewords or streams. Therefore, there is no user data exchange needed in F-NCJT. Also, since all coordinated TRPs need to synchronize scheduling information, a centralized scheduler is needed. The received signal F-NCJT can be expressed as
\begin{align}
\mathbf{y}_k=\mathbf{H}_k\begin{bmatrix}
\mathbf{W}_{k,1} &  &  & \\ 
 & \mathbf{W}_{k,2} &  & \\ 
 &  & \ddots & \\ 
 &  &  & \mathbf{W}_{k,|\mathcal{C}_b|}
\end{bmatrix}\mathbf{s}_k+\mathbf{I}_o+\mathbf{n}
\label{equ_FNCJT}
\end{align}
where the $N_T\times 1 $ matrix $\mathbf{W}_{k,l}$ denotes the $k$th user's precoding matrix from the $l$th coordinated TRP in $\mathcal{C}_b$, and $\mathbf{s}_k$ is the $|\mathcal{C}_b|\times 1$ symbol vector of $k$th user's. It can be observed that the aggregate precoding matrix is a block diagonal matrix, and its off-diagonal elements are zeros. 

The scheduling algorithm of F-NCJT is described in Fig. \ref{fig_F_NCJT}. The scheduling procedure is similar to DPS but with two key differences. One is that the sum spectral efficiency of each UE across all TRPs is calculated as inputs to the centralized scheduler. The other one is that user data does not need to be shared among all coordinated TRPs in a CoMP set and only CSI reports need to be exchanged, which relaxes the requirement on backhaul links.

\begin{figure}
\hrulefill
\begin{algorithmic}[1]
\For{$b=1,2,\cdots, N_b$}
\State User vector $U_b$ of the $b$th TRP
\State Identify the CoMP set $\mathcal{C}_b$ of the $b$th TRP
\State Read the indices $\left\lbrace b_1,b_2,\cdots,b_{|\mathcal{C}_b|}\right\rbrace$ of TRPs in $\mathcal{C}_b$
\For{$i=1,2,\cdots,|\mathcal{C}_b|$}
\State $U_b=U_b\bigcup U_{b_i}$
\EndFor
\If{$\mathcal{C}_b$ is scheduled}
\State The $b$th TRP reads PRB allocations from $\mathcal{C}_b$
\Else
\For {$j=1,2,\cdots,|U_b|$}
\State Compute aggregate spectral efficiency $\eta_j=\sum\limits_{i=1}^{|\mathcal{C}_b| }c_{j,b_i}$
\EndFor
\State Perform proportional fair scheduling for $\mathcal{C}_b$
\State $\mathcal{C}_b$ is scheduled = true
\EndIf
\EndFor
\end{algorithmic}
\hrulefill
\caption{Scheduling algorithm for F-NCJT.}
\label{fig_F_NCJT}
\end{figure}

The main gain of F-NCJT is that the UE is able to estimate channel matrices from coordinated TRPs and then perform MMSE-IRC receiver to suppress interlayer interference.

\subsection{NF-NCJT}
NF-NCJT works in a distributed manner, where each coordinated TRP schedules users independently via a local scheduler. There is no user data or CSI report exchange needed in NF-NCJT. Assuming the first $l$ coordinated TRPs in the CoMP set schedule the UE in the same PRB,
the received signal of NF-NCJT can be expressed as
\begin{align}
&\mathbf{y}_k\nonumber\\
&=\mathbf{H}_k\begin{bmatrix}
\mathbf{W}_{k,1} &  &  & & & \\ 
 & \ddots &  & & & \\   
 &  & \mathbf{W}_{k,l} & & & \\ 
 &  &  &\mathbf{W}'_{k,l+1} & & \\  
 &  &  & &\ddots & \\ 
 &  &  &  & & \mathbf{W}'_{k,|\mathcal{C}_b|}
\end{bmatrix}\left[\mathbf{s}_k ; \mathbf{s}'_k \right] \nonumber\\
&+\mathbf{I}_o+\mathbf{n}
\end{align}
where $\mathbf{W}_{k,1},\cdots,\mathbf{W}_{k,l}$ are precoding matrices for the UE, $\mathbf{W}_{k,l+1},\cdots,\mathbf{W}_{k,|\mathcal{C}_b|}$ are precoding matrices of interference, $\mathbf{s}_k$ is the $l\times 1$ signal vector for the UE, and $\mathbf{s}'_k$ is the $(|\mathcal{C}_b|-l)\times 1$ interference signal vector. As local schedulers do not synchronize, there is a chance that a UE is scheduled on the same PRB across all coordinated TRPs in the CoMP set. When this happens, i.e., $l=|\mathcal{C}_b|$, the received signal of NF-NCJT coincides with (\ref{equ_FNCJT}).

The scheduling algorithm of NF-NCJT is described in Fig. \ref{fig_NF_NCJT}. First, all UEs within a CoMP set need to be added to the user vector of each coordinated TRP. Second, the spectral efficiency of each UE in each TRP is calculated as inputs to the distributed scheduler. The distributed scheduler then allocates resources to UEs according to the proportional fair rule. The whole scheduling procedure needs to be done on a per TRP basis. As a result, neither user data nor CSI reports need to be shared among coordinated TRPs in a CoMP set.  

\begin{figure}
\hrulefill
\begin{algorithmic}[1]
\For{$b=1,2,\cdots, N_b$}
\State User vector $U_b$ of the $b$th TRP
\State Identify the CoMP set $\mathcal{C}_b$ of the $b$th TRP
\State Read the indices $\left\lbrace b_1,b_2,\cdots,b_{|\mathcal{C}_b|}\right\rbrace$ of TRPs in $\mathcal{C}_b$
\For{$i=1,2,\cdots,|\mathcal{C}_b|$}
\State $U_b=U_b\bigcup U_{b_i}$
\EndFor

\For {$j=1,2,\cdots,|U_b|$}
\State Compute spectral efficiency $c_{j,b}$
\EndFor
\State Perform proportional fair scheduling for the $b$th TRP

\EndFor
\end{algorithmic}
\hrulefill
\caption{Scheduling algorithm for NF-NCJT.}
\label{fig_NF_NCJT}
\end{figure}

The main gain of NF-NCJT is from the degrees of freedom of scheduling in each coordinated TRP.

\section{Results and Analysis} \label{sec_results_and_analysis}

This section studies the performance of the abovementioned network coordination schemes via system level simulation. The result of no network coordination is shown as reference. Typical simulation assumptions are used for performance evaluation and are listed in Table \ref{tab_sim_setting}.

\begin{table}[ht]
\caption{Key simulation assumptions.}
\center
\scriptsize
    \begin{tabular}{|c|c|}
    \hline
    Parameter  & Value  \\ \hline
Layout  & Indoor scenario with 8 TRPs \cite{36741}  \\ \hline        
        Inter-side distance  & $30$ m  \\ \hline
        
        Carrier frequency  & $3.5$ GHz  \\ \hline
        Bandwidth & $10$ MHz  \\ \hline
                Subcarrier spacing & $15$ kHz  \\ \hline
                Channel model &  3GPP TR 36.814 Indoor hotspot \cite{36814}  \\ \hline
                TRP antenna configuration & ULA with $2$ elements  \\ \hline

                TRP transmit power & $24$ dBm \\ \hline
                TRP antenna pattern & Omni-directional with 5 dBi gain  \\ \hline
                                TRP antenna height & 6 m  \\ \hline

                UE antenna heigh & 1.5 m   \\ \hline
                UE dropping &  uniform   \\ \hline
                                  UE antenna & $N_R=4$ \\ \hline

                 UE antenna gain & 0 dB \\ \hline
                                 Traffic  model & FTP with 0.5 Mbytes file and 10/s arrival rate size \\ \hline
                                                                   UE receiver & MMSE-IRC \\ \hline
                                                                   Feedback delay & 5 ms \\ \hline
                                                                   Transmission mode & LTE TM 10 \\ \hline
                                                                                                                                      Channel estimation & Realistic \\ \hline

    \end{tabular}
    \label{tab_sim_setting}
\end{table}

\begin{figure}[t]
\centering
\includegraphics[width=3.5in]{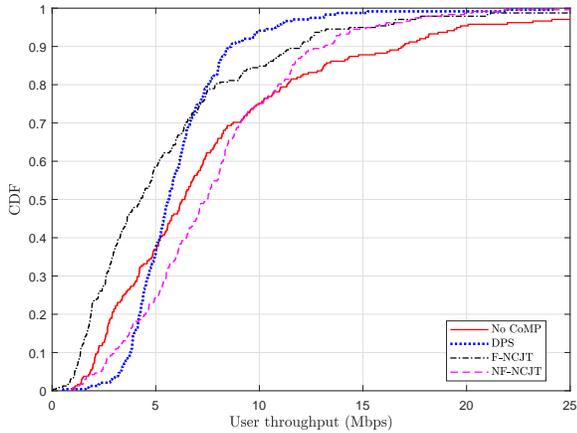}    
\caption{CDFs of user throughputs of different network coordination schemes (3 users per TRP).}
\label{fig_Comparison_Nuser_3_GCWS}
\end{figure}

Fig. \ref{fig_Comparison_Nuser_3_GCWS} depicts cumulative distribution functions (CDFs) of user throughputs of different network coordination schemes with 3 users per TRP. It can be observed that F-NCJT has the worst cell-edge and median user throughputs because the lack of scheduling freedom. A cell-center F-NCJT user, for instance, is expected to have a strong link from the closest TRP and a number of poor links from other coordinated TRPs. The signal power of these poor links is so small that their corresponding resources have a high probability of being wasted. On the other hand, a cell-edge F-NCJT user will be connected to have multiple relatively weak links. This cell-edge F-NCJT user should use all of its spatial resources to concentrate on receiving one layer from one TRP to guarantee certain throughput. However, by the definition of F-NCJT, the cell-edge user still tries to receive multiple layers from coordinated TRPs, which significantly reduces user throughput. DPS has the best performance in cell edge (5th percentile user throughput), since other coordinated TRPs are muted to minimize interference. However, this results in lower resource efficiency. As a result, DPS has the lowest cell-center user throughput (95th percentile user throughput) and the median user throughput DPS is lower than that of NF-NCJT and no network coordination. NF-NCJT improves the median user throughput, although it has a worse cell-edge user throughput than DPS and a worse cell-center user throughput than no network coordination. 

\begin{figure}[t]
\centering
\includegraphics[width=3.5in]{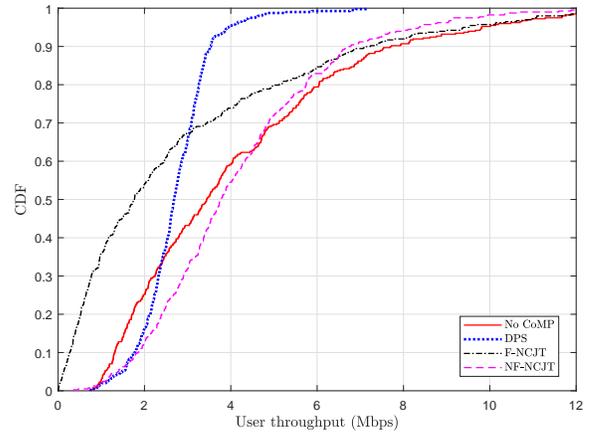}    
\caption{CDFs of user throughputs of different network coordination schemes (5 users per TRP).}
\label{fig_Comparison_Nuser_5_GCWS}
\end{figure}

CDFs of user throughputs of different network coordination schemes with 5 users per TRP are illustrated in Fig. \ref{fig_Comparison_Nuser_5_GCWS}. As the traffic becomes richer, the gap between median user throughputs of F-NCJT and other network coordination schemes enlarges, and the gain of DPS over NF-NCJT at cell edge drops. The performance of NF-NCJT is relatively less sensitive to traffic density in the network that other network coordination schemes.   

\begin{figure}[t]
\centering
\includegraphics[width=3.5in]{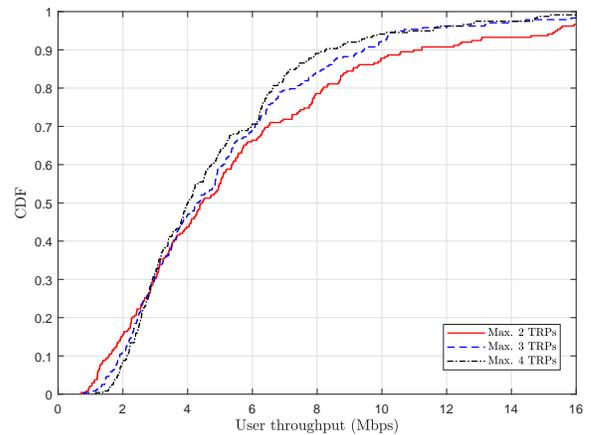}    
\caption{Performance comparison between different maximum numbers of  coordinated TRPs in NF-NCJT.}
\label{fig_m2m3m4_GCWS}
\end{figure}

It is interesting to study the optimal maximum number of coordinated TRPs in NF-NCJT as this is related to   resource efficiency and UE efficiency. Fig. \ref{fig_m2m3m4_GCWS} compares the user throughputs of different maximum numbers of coordinated TRPs in NF-NCJT. It can be observed that at cell edge (5th percentile user throughput), user throughput increases with the maximum number of coordinated TRPs. Let maximum equals 2 be the reference. When the maximum number of coordinated TRPs equals three, the gain at the 5th percentile user throughput is 30\%. This gain enlarges to 50\% when the maximum number of coordinated TRPs equals four. This performance improvement is achieved through allowing cell edge UEs to have more degrees of freedom to choose TRPs with better link qualities. However, increasing the maximum number of coordinated TRPs in NF-NCJT will at the same time significantly decrease the 95th percentile user throughput. Approximately 30\% drop at the 95th percentile user throughput is seen when maximum equals either three or four. This loss is because more resources are occupied by cell edge UEs from both own cell and coordinating cells. Also, median user throughputs drop by 10\% to 15\% when maxim equals three or four.

\section{Conclusions} \label{conclusion_section}
Flexible network coordination is expected to play an important role in NR. For the two centralized methods, DPS performs well in cell edge because of blanking PRBs from coordinated TRPs. However, it requires sharing of both user data and CSI reports. Although F-NCJT does not require exchange of user data, its performance is largely limited to the lack of freedom during scheduling, resulting in the worse performance among the three. On the contrary, NF-NCJT operates in a distributed manner, without the need of sharing user data or CSI report. The freedom in scheduling yields scheduling gain, which provides the highest median user throughput. Distributed scheduling brings an reasonably efficient and low-cost solution to network coordination. Therefore, the deployment of DRAN architecture should be considered in 5G networks. For future work, it is essential to study flexible transition among different network coordination schemes, including CJT. Additionally, performance evaluations can be conducted in other scenarios.

\section*{Acknowledgment}
Part of this work has been performed in the framework of the Horizon 2020 project ONE5G (ICT-760809) receiving funds from the European Union. The authors would like to acknowledge the contributions of their colleagues in the project, although the views expressed in this contribution are those of the authors and do not necessarily represent the project.

\end{document}